\documentclass[10pt,twocolumn,letterpaper]{article}

\usepackage{iccv}              
\usepackage{multirow}
\usepackage[accsupp]{axessibility}  

\definecolor{gold}{HTML}{FAE37F}
\definecolor{silver}{HTML}{D7D7D7}
\definecolor{bronze}{HTML}{EDBA91}
\newcommand{\au}[1]{\setlength{\fboxsep}{1pt}\colorbox{gold}{#1}}
\newcommand{\ag}[1]{\setlength{\fboxsep}{1pt}\colorbox{silver}{#1}}
\newcommand{\cu}[1]{\setlength{\fboxsep}{1pt}\colorbox{bronze}{#1}}

\definecolor{iccvblue}{rgb}{0.21,0.49,0.74}
\usepackage[pagebackref,breaklinks,colorlinks,allcolors=iccvblue]{hyperref}

\def\paperID{3821} 
\def\confName{ICCV}
\def\confYear{2025}

\title{\includegraphics[scale=0.025, keepaspectratio]{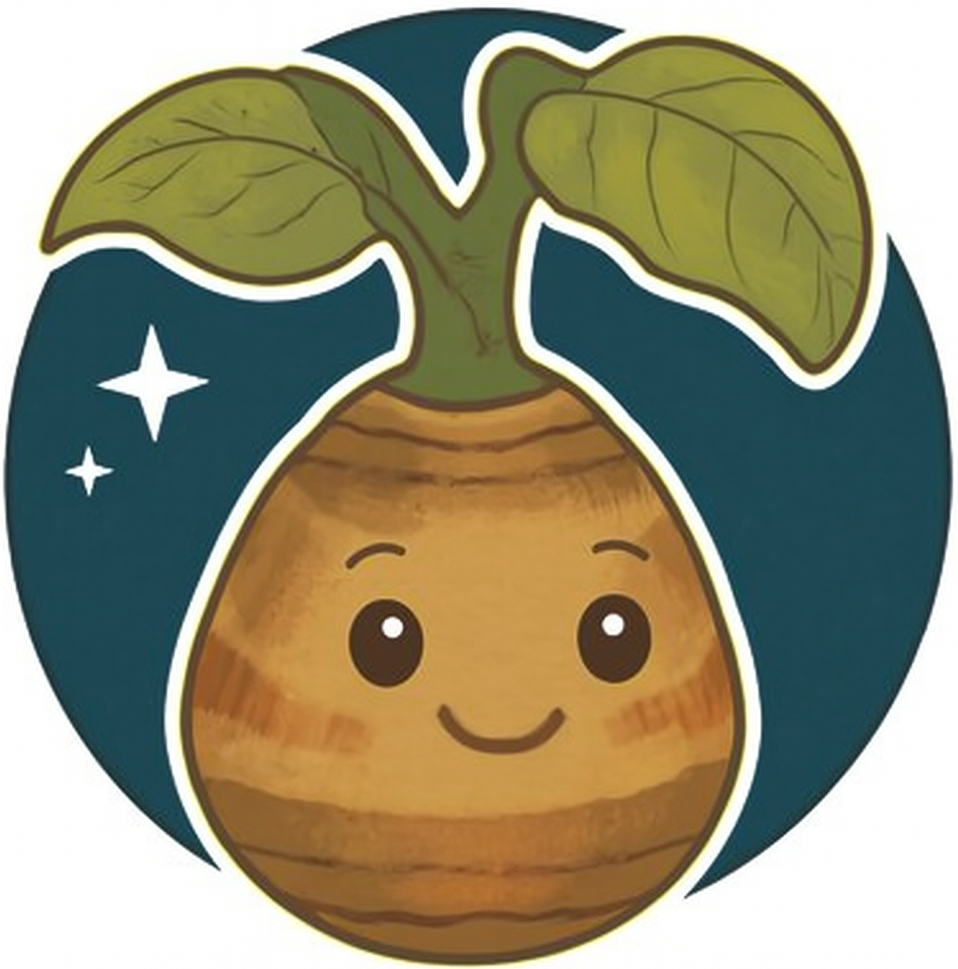} TARO: Timestep-Adaptive Representation Alignment \\ with Onset-Aware Conditioning for Synchronized Video-to-Audio Synthesis}

\author{Tri Ton \quad Ji Woo Hong \quad Chang D. Yoo \\
Korea Advanced Institute of Science and Technology (KAIST) \\
{\tt\small \{tritth, jiwoohong93, cd\_yoo\}@kaist.ac.kr}
}

\begin{document}


\maketitle

\begin{abstract}
This paper introduces Timestep-Adaptive Representation Alignment with Onset-Aware Conditioning (TARO), a novel framework for high-fidelity and temporally coherent video-to-audio synthesis. Built upon flow-based transformers, which offer stable training and continuous transformations for enhanced synchronization and audio quality, TARO introduces two key innovations: (1) Timestep-Adaptive Representation Alignment (TRA), which dynamically aligns latent representations by adjusting alignment strength based on the noise schedule, ensuring smooth evolution and improved fidelity, and (2) Onset-Aware Conditioning (OAC), which integrates onset cues that serve as sharp event-driven markers of audio-relevant visual moments to enhance synchronization with dynamic visual events. Extensive experiments on the VGGSound and Landscape datasets demonstrate that TARO outperforms prior methods, achieving relatively 53\% lower Frechet Distance (FD), 29\% lower Frechet Audio Distance (FAD), and a 97.19\% Alignment Accuracy, highlighting its superior audio quality and synchronization precision.
The code is available at: {\href{https://github.com/triton99/TARO}{\nolinkurl{github.com/triton99/TARO}.}}
\end{abstract}    
\section{Introduction}
\label{sec:intro}

Video-to-audio generation has garnered increasing attention due to its broad applications in content creation and film production \cite{ament2014foley, choi2022proposal}. The goal is to synthesize audio that is both high-quality and temporally aligned with video frames. However, achieving high-fidelity and precisely synchronized audio remains challenging, especially when dealing with rapidly changing visual content.

Despite progress in video-to-audio generation, existing approaches struggle to balance fidelity, synchronization, and efficiency. GAN-based methods \cite{chen2020generating} improve audio fidelity but often fail to maintain temporal coherence. Transformer-based autoregressive methods \cite{iashin2021taming, sheffer2023hear} offer greater modeling expressiveness but suffer from slow inference and accumulating timing errors. Diffusion-based methods \cite{luo2024diff, xing2024seeing, zhang2024foleycrafter, pham2024mdsgen} improve fidelity through latent diffusion and cross-modal pretraining, yet their high computational costs hinder real-time deployment. In contrast, flow-based methods \cite{esser2024scaling, yang2024cogvideox} provide stable training and continuous transformations, emerging as a promising alternative for efficient video-to-audio synthesis.

\begin{figure}
    \centering
    \includegraphics[width=\linewidth,keepaspectratio]{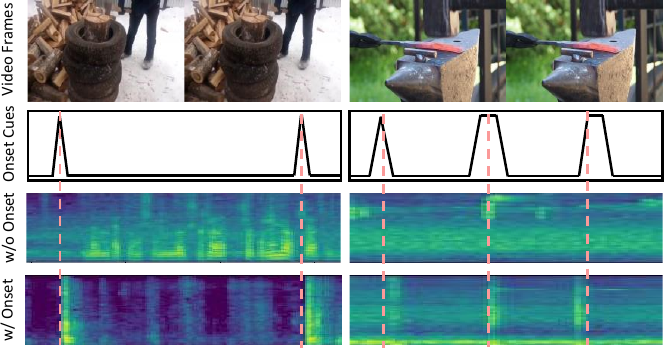}
    \caption{\textbf{Our TARO leverages onset-aware conditioning to improve synchronization, aligning generated audio with event-driven cues in video.}}
    \label{fig:quantitative_1}
    \vspace{-0.5cm}
\end{figure}

Existing video-to-audio approaches naively learn latent audio representation through reconstruction loss, which can lead to suboptimal quality. Recently, REPA \cite{yu2024representation} has attempted to leverage pretrained features \cite{oquab2023dinov2} as a constraint to guide latent representations in image synthesis. Inspired by this, we explore applying a similar constraint-driven approach to latent audio representation learning.
Directly applying methods like REPA is non-trivial, so we carefully design TRA by (1) dynamically adjusting the alignment strength based on the diffusion noise schedule, (2) addressing sequence length mismatch, and (3) selecting appropriate audio encoder outputs. This prevents over-regularization while preserving semantic grounding throughout training.
In particular, adapting this regularization for video-to-audio synthesis is challenging, as imposing strict constraints on latent representations in early timesteps, when noise dominates, can hinder learning rather than improve alignment. Instead, a more adaptive strategy is required to gradually refine alignment as the latent representation evolves.
Additionally, achieving precise synchronization at the event level remains an open challenge. Existing methods struggle to capture fine-grained auditory cues tied to specific visual events, often generating misaligned or unnatural sounds. This issue arises from the lack of explicit mechanisms to detect and condition on key auditory events such as footsteps, object interactions, or percussive impacts, which serve as strong temporal anchors for realistic audio synthesis.

\begin{figure}
    \centering
    \includegraphics[width=\linewidth,keepaspectratio]{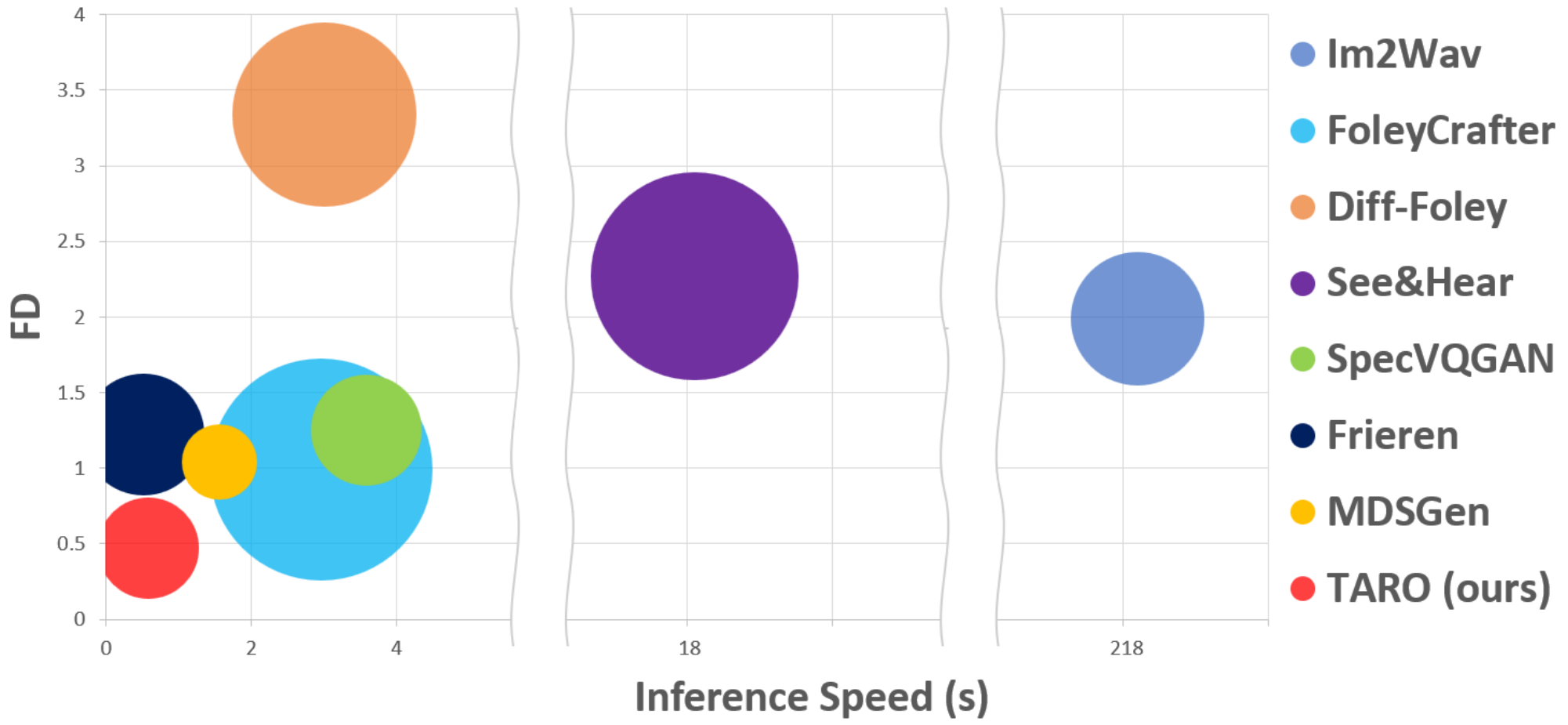}
    \caption{\textbf{Comparison of FD (y-axis) and inference speed (x-axis) across different models, with marker size representing parameter size.}
    }
    \label{fig:efficiency}
    \vspace{-0.5cm}
\end{figure}

To address these challenges, we introduce Timestep-Adaptive Representation Alignment with Onset-Aware Conditioning (TARO), a novel flow-based transformer framework that enhances both fidelity and synchronization in video-to-audio generation.
TARO introduces two key innovations to improve latent audio representation and event-level alignment.
First, Timestep-Adaptive Representation Alignment (TRA) dynamically adjusts alignment strength based on the noise schedule, ensuring progressive refinement of latent representations during training.
Second, Onset-Aware Conditioning (OAC) explicitly integrates event-driven cues to align generated audio with video motion by detecting key auditory events. These sparse and semantically meaningful onset cues serve as temporal anchors, guiding the model to generate sound at appropriate moments. This conditioning mechanism enhances temporal precision, improving both perceptual quality and the naturalness of video-to-audio synthesis.
Extensive experiments on two public datasets, VGGSound \cite{chen2020vggsound}, and Landscape \cite{lee2022sound} datasets, demonstrate that TARO synthesizes high-quality, well-aligned audio, outperforms state-of-the-art (SOTA) methods across various comprehensive metrics
as shown in Figure \ref{fig:efficiency}.

In summary, our contributions are as follows:

\begin{itemize}
\item We propose
\textbf{T}imestep-\textbf{A}daptive \textbf{R}epresentation Alignment with \textbf{O}nset-Aware Conditioning (TARO), a novel flow-based transformer framework for video-to-audio generation. TARO integrates both visual and onset-based control signals to enhance the fidelity and synchronization of generated audio.
\item We introduce a novel Timestep-Adaptive Representation Alignment strategy that dynamically adjusts alignment strength based on the noise schedule, refining latent representations for high-quality, temporally coherent audio. 
\item We present an Onset-Aware Conditioning that captures key event-driven cues to refine synchronization, enabling precise audio alignment with dynamic visual events.
\end{itemize}

\section{Related Works}
\label{relatedworks}

\noindent \textbf{Video-to-audio generation.}
Video-to-audio generation, a core task in multimodal generation \cite{ruan2023mm, luu2025enhancing}, aims to synthesize temporally synchronized and semantically relevant audio from video.
Transformer-based autoregressive methods like SpecVQGAN \cite{iashin2021taming} offer flexibility but suffer from slow inference and weak alignment. 
Im2Wav \cite{sheffer2023hear} extends this to images via CLIP \cite{radford2021learning} embeddings but struggles with dynamic video content. Diffusion-based models \cite{ho2020denoising, song2020denoising, yoon2024frag, yoon2024dni, yoon2024tpc, hong2025ita} improve quality but require multiple steps and large parameters.
Diff-Foley \cite{luo2024diff} enhances alignment via latent diffusion, while See \& Hear \cite{xing2024seeing} improves multimodal integration but lacks fine-grained temporal precision.
FoleyCrafter \cite{zhang2024foleycrafter} incorporates a temporal controller for high-quality Foley sound but remains computationally heavy. 
Frieren \cite{wang2025frieren} reduces steps via rectified flow matching but may miss sparse, event-driven dynamics. 
MDSGen \cite{pham2024mdsgen} employs masked diffusion for efficiency but still relies on multiple steps for fidelity. 
Furthermore, several works incorporate auxiliary temporal features as condition information to guide the generation process. TiVA \cite{wang2024tiva} leverages smoothed energy control, and ReWaS \cite{jeong2025read} uses down-sampled mel-spectrograms. However, these signals often introduce temporal redundancy or smoothing artifacts, making them less effective at capturing fine-grained event boundaries critical for precise alignment.
Our TARO framework addresses these issues using 
timestep-adaptive alignment, and onset-aware conditioning to enhance audio quality, synchronization, and computational efficiency.

\noindent \textbf{Flow-based generative models.}
Flow-based generative models have gained traction for their efficiency and stability in image and audio synthesis, providing an alternative to score-based diffusion models. Flow matching \cite{lipman2022flow} models a continuous vector field to smoothly transport samples from noise to data using ordinary differential equations (ODEs), achieving robust training and superior performance.
In audio synthesis, Voicebox \cite{le2023voicebox} and Audiobox \cite{vyas2023audiobox} leverage flow matching for speech and unified audio generation, while VoiceFlow \cite{guo2024voiceflow} applies rectified flow matching for efficient text-to-speech. 
Our work bridges this gap by extending flow-based transformers to video-to-audio synthesis to improve generation quality and synchronization.

\noindent \textbf{Audio Representation Learning.}
Recent advances in representation learning \cite{he2022masked, luu2024predictive} have significantly improved the ability to extract meaningful features for audio understanding, offering valuable insights for generative models.
Self-supervised learning frameworks such as wav2vec 2.0 \cite{baevski2020wav2vec} capture robust speech representations, while CLAP \cite{wu2023large} employs contrastive language-audio pretraining to learn general-purpose audio semantics. Transformer-based models like BEATs \cite{chen2022beats} leverage masked spectrogram learning for audio classification, whereas EAT \cite{chen2024eat} introduces an efficient audio transformer trained with utterance-frame objectives.
Despite their effectiveness, these models have yet to be fully leveraged for video-to-audio tasks, where precise temporal alignment and semantic consistency are critical. Our approach integrates pretrained audio encoders into a timestep-adaptive alignment mechanism, ensuring that latent representations evolve meaningfully across synthesis.

\section{Preliminary}
\label{sec:preliminary}

Flow-based generative models provide an alternative to Denoising Diffusion Probabilistic Models (DDPM) \cite{ho2020denoising} by leveraging a continuous-time framework for data generation. Unlike DDPM, which employs discrete noise schedules, flow-based models \cite{lipman2022flow, ma2024sit} define a smooth perturbation process over $t \in [0, 1]$, continuously evolving data from a noise distribution to a structured target distribution. To model this transformation, given data $x_* \sim p(x|c)$, conditioned on $c$, a perturbation process is defined as:
\begin{equation}
    x_t = \alpha_t x_* + \sigma_t \epsilon,
\end{equation}
where $\alpha_t$ is a decreasing function of $t$ for controlling the data component and $\sigma_t$ is an increasing function of $t$ for governing the noise contribution, and Gaussian noise $\epsilon \sim \mathcal{N}(0, I)$. This process follows a probability flow described by an ordinary differential equation (ODE):
\begin{equation}
    \dot{x_t} = v(x_t, t, c),
\end{equation}
where $v(x_t, t, c)$ represents the transport vector field guiding the flow from noise to data, conditioned on $c$.

To enable efficient sample generation, this ODE is numerically solved, such as using the Euler method, starting from random noise $\epsilon \sim \mathcal{N}(0, I)$. The transport vector field is parameterized as $v_\theta(x_t, t, c)$, typically implemented as a neural network, and trained using the conditional flow matching (CFM) objective:
\begin{equation}
 \mathcal{L}_{\text{CFM}} = \mathbb{E}_{t, (x_*, c),\epsilon} \left[ \| v_\theta(x_t, t, c) - (\dot{\alpha}_t x_* + \dot{\sigma}_t \epsilon) \|_2^2 \right],
\end{equation}
where $\dot{\alpha}_t x^* + \dot{\sigma}_t \epsilon$ represents the target velocity. 

Our approach follows the stochastic interpolant framework proposed by \cite{ma2024sit}, which unifies flow-based and diffusion models under a flexible interpolant design. We adopt a linear interpolant, defined as $\alpha_t = 1 - t$ and $\sigma_t = t$, which provides a simple yet effective mapping from noise to data over $t \in [0, 1]$.
This choice ensures stable training while preserving temporal coherence in audio generation.
Additionally, our approach is adaptable to other interpolants, such as variance-preserving (VP) interpolants, defined by $\alpha_t = cos(\frac{\pi}{2}t)$ and $\sigma_t = sin(\frac{\pi}{2}t)$, which offer smoother noise schedules.

\section{Method}
\label{method}

\subsection{Overview of TARO}

\begin{figure*}
    \centering
    \includegraphics[width=0.95\linewidth,keepaspectratio]{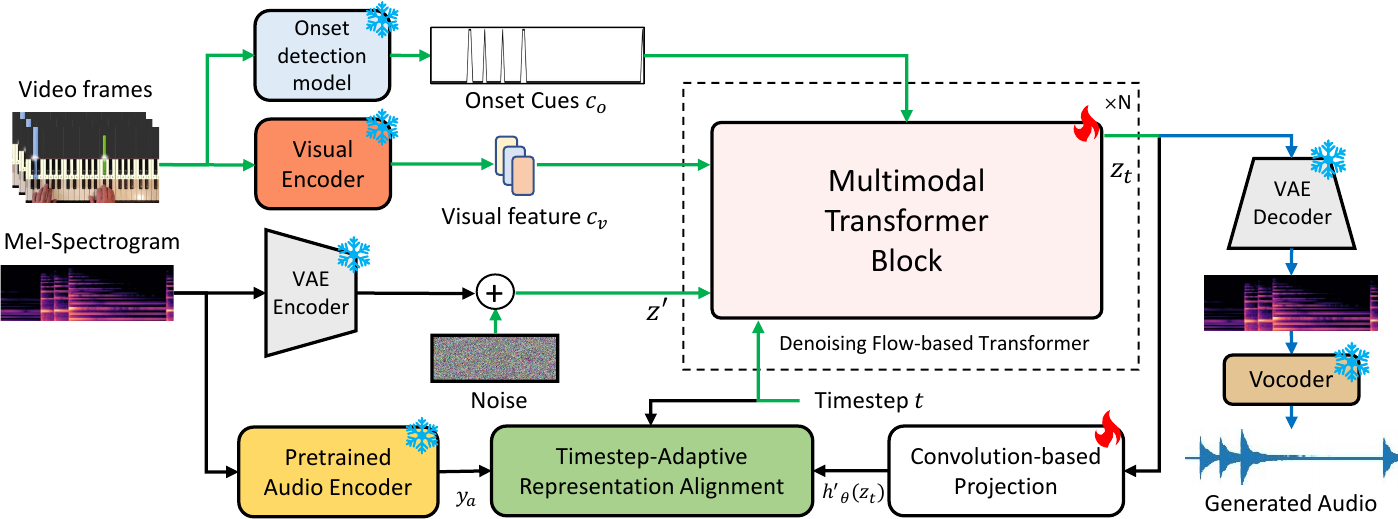}
    \caption{\textbf{Overview of TARO.} Our TARO is a flow-based multimodal transformer for video-to-audio generation, integrating Timestep-Adaptive Representation Alignment (TRA) and Onset-Aware Conditioning (OAC) to enhance synchronization and fidelity.
    Black arrows $\rightarrow$ denote branches used only in training, blue arrows $\color{blue}\rightarrow$ for inference only, and green arrows $\color{green}\rightarrow$ for both training and inference.}
    \label{fig:framework}
    \vspace{-0.5cm}
\end{figure*}

Our approach builds upon the MMDiT \cite{esser2024scaling}, adapting it for video-to-audio synthesis by incorporating event-driven synchronization and representation alignment mechanisms.
As illustrated in Figure \ref{fig:framework}, the pipeline begins with processing the input video through a pretrained visual encoder \cite{luo2024diff} with frozen parameters to extract frame-level features, denoted as $c_v \in \mathbb{R}^{L_v \times D_v}$ where $L_v$ and $D_v$ represent the number of frames and the visual feature dimension, respectively. These features capture the overall scene structure and serve as the primary conditioning input for audio synthesis.
To enhance audio-visual synchronization, we introduce an Onset-Aware Conditioning (OAC), which employs a pretrained onset detection model \cite{comunita2024syncfusion} to generate onset cues $c_o \in \mathbb{R}^{1\times L_o}$, identifying key moments where audio-relevant events occur in the video frames.

During training, a mel-spectrogram $M \in \mathbb{R}^{F \times T}$, where $F$ represents the frequency bins and $T$ denotes the temporal resolution, is encoded through a Variational Autoencoder (VAE) to produce a latent representation $z \in \mathbb{R}^{8 \times F/r \times T/r}$, where $r$ is the downsampling factor. This latent representation is further patched and tokenized with a patch size of $p=2$, yielding $z' \in \mathbb{R}^{L_z \times D_z}$, where $L_z$ represents the sequence length of the tokens, and $D_z$ is the hidden dimension. These tokenized latents are then processed by a flow-based multimodal transformer, which consists of $N$ stacked multimodal transformer blocks. Throughout this process, the model integrates cross-modal attention by conditioning on visual features $c_v$ and onset cues $c_o$, ensuring that the generated audio aligns with the video's structural and event-driven information.

To enable effective multimodal integration, our framework concatenates the projected visual features and audio latent embeddings along the sequence dimension before feeding them into the transformer blocks, as illustrated in Figure \ref{fig:mmditblock}. Given the differences in feature spaces and numerical scales between these modalities, Adaptive Layer Normalization (AdaLN) is applied separately to each modality, ensuring consistent processing and preventing imbalance between visual and audio representations. These processed tokens then interact through a joint attention mechanism, where query, key, and value representations from both modalities are concatenated and jointly attended, allowing the model to capture fine-grained audio-visual correspondences and refine the latent representation.
A key component of our framework is the Timestep-Adaptive Representation Alignment (TRA), which dynamically aligns latent representations with high-quality pretrained audio encoders \cite{chen2024eat, chen2022beats, wu2023large, baevski2020wav2vec}. 
This adaptive strategy leverages pretrained audio priors to enhance both audio fidelity and synchronization with visual content.

During inference, the model initializes the latent audio representation from Gaussian noise and iteratively refines it through the flow-based denoising process, guided by visual features $c_v$ and onset cues $c_o$. The final latent is decoded by the VAE decoder to reconstruct the mel-spectrogram $\hat{M} \in \mathbb{R}^{F \times T}$, which is then converted into an audio waveform using a vocoder.
The following sections detail the OAC in Sec. \ref{method:onset_module}, followed by the TRA in Sec. \ref{method:time_repa}.

\subsection{Onset-Aware Conditioning}
\label{method:onset_module}

Accurate audio-video synchronization requires identifying event-driven cues that indicate key auditory moments. Onset cues capture abrupt sound changes, such as percussive strikes, footsteps, or sudden object interactions, providing structured timing signals crucial for aligning acoustic events with visual motion. While pretrained visual features \cite{luo2024diff} encode semantic content and broad temporal correlations, they lack explicit event timing. In contrast, onset cues introduce discrete temporal markers, refining fine-grained synchronization in video-to-audio generation.
To incorporate these cues, we utilize a pretrained video onset detection model \cite{comunita2024syncfusion} to extract frame-wise onset events $c_o$, which highlight key transitions in the video. As illustrated in Figure \ref{fig:mmditblock}, these cues undergo a projection module using multi-layer perceptions (MLPs) to ensure alignment with the latent space. To reinforce temporal coherence, the projected onset features are summed with timestep embeddings before integration into the network. This step allows the model to jointly encode both event timing and global progression through the generative process, ensuring more structured conditioning.
We employ AdaLN to modulate both visual features and audio latents, serving as guidance signals in a lightweight and temporally aligned manner, enabling fine-grained adaptation across different video frames. AdaLN dynamically scales and shifts feature activations based on the onset-conditioned parameters, ensuring flexible processing throughout the generative process. 
This design avoids the computational overhead of token-level concatenation or cross-attention, which is commonly used in DiT-style methods. 
Specifically, given an input representation $x$ and and an onset-based conditioning signal $c$, AdaLN is computed as:
\begin{equation}
    \text{AdaLN}(x | \gamma_{c}, \beta_{c}) = \gamma_{c} x + \beta_{c},
\end{equation}
where modulation parameters are obtained by MLP:
\begin{equation}
    \gamma_{c} = \text{MLP}_\gamma(c), \quad \beta_{c} = \text{MLP}_\beta(c).
\end{equation}

By integrating onset cues with timestep embeddings, the model captures both fine-grained event timing and the overall generative progression, ensuring stable synchronization throughout the synthesis process. 

\subsection{Time-Adaptive Representation Alignment} 
\label{method:time_repa}

The standard alignment method \cite{yu2024representation} applies fixed constraints throughout synthesis, which can hinder learning in early timesteps when noise dominates.
TRA dynamically adjusts alignment strength based on the noise schedule, allowing latent representations to evolve progressively.
This adaptive approach prevents overly strict constraints in noisy stages while gradually reinforcing structure as synthesis progresses, ensuring smooth and effective alignment.
Given a pretrained encoder $f_{enc}$ and a clean audio sample $x_a$, we compute its embedding as $y_a = f_{enc}(x_a) \in \mathbb{R}^{L_a \times D_a}$, where $L_a$ represents the number of patches and $D_a$ denotes the embedding dimension. To align with this reference, we use a trainable projection head $h_\phi$, which transforms the model latent representation $z_t \in \mathbb{R}^{L_z \times D_z}$ at timestep $t$ into the same $y_a$ feature space as $h_\phi(z_t) \in \mathbb{R}^{L_z \times D_a}$. This projection ensures that the learned representation matches the structured features of clean audio, enabling effective alignment.
To achieve this alignment, we introduce a time-weighted alignment loss, adjusting alignment strength dynamically based on the noise schedule:
\begin{equation}
\mathcal{L}_{\text{align}} = \sum_{t} w(t) \cdot d(h'_{\phi}(z_t), y_a),
\end{equation}
where $d(\cdot, \cdot)$ is cosine similarity. The weighting function $w(t)$ is computed as:

\begin{figure}[t]
    \centering
    \includegraphics[width=\linewidth,keepaspectratio]{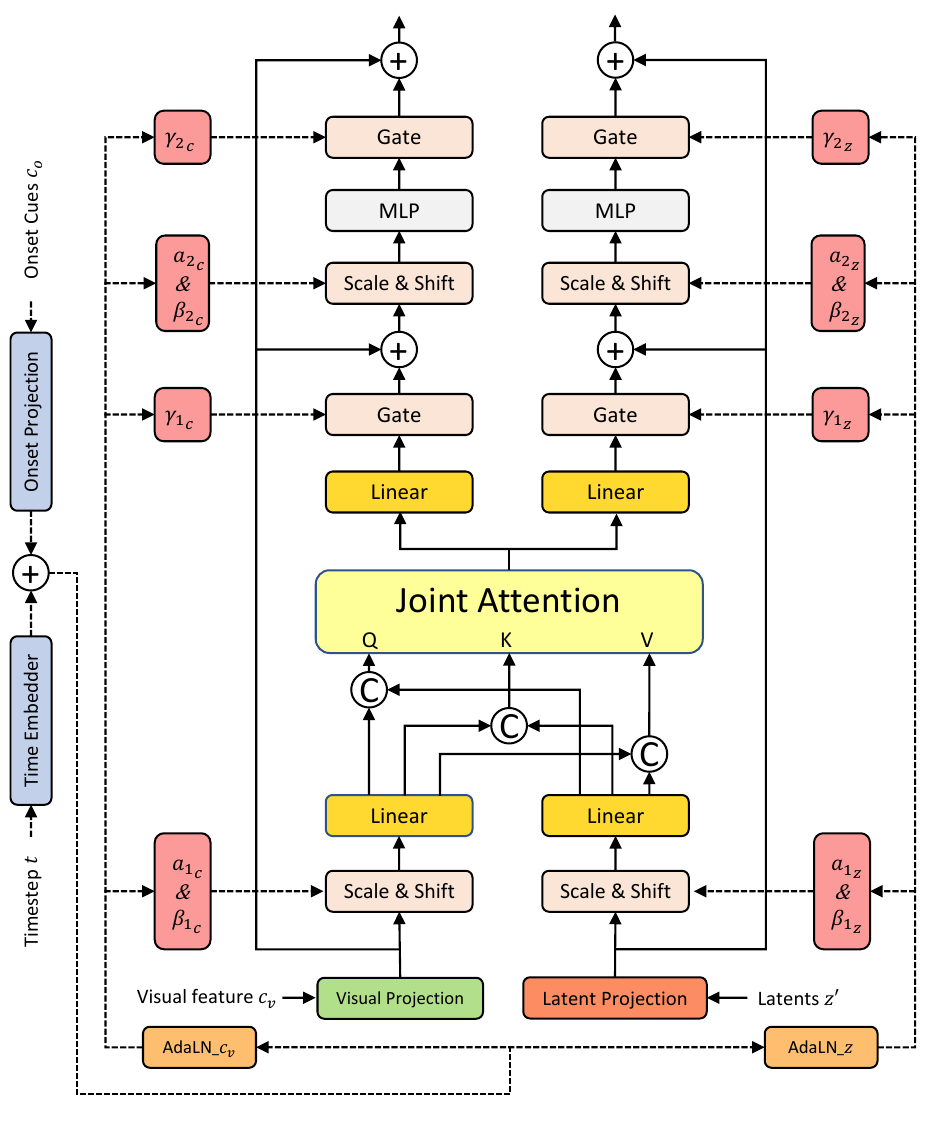}
    \vspace{-0.8cm}
    \caption{\textbf{One Multimodal Transformer Blocks}. The block integrates onset cues, visual features, and latent audio representations through adaptive modulation and joint attention.$+$ denotes summation, and $\text{C}$ represents concatenation.}
    \label{fig:mmditblock}
    \vspace{-0.5cm}
\end{figure}

\begin{equation}
w(t) = \text{sigmoid}(\text{log}(\frac{\alpha_t}{\sigma_t + \epsilon})),
\end{equation}
where $\alpha_t$ and $\sigma_t$ are interpolant coefficients from the noise schedule, and $\epsilon$ constant prevents division by zero. This log-sigmoid formulation smoothly scales alignment strength, ensuring weaker constraints in early timesteps when noise dominates and stronger constraints as the latent representations converge toward structured audio features. The total loss is given by:
\begin{equation}
    \mathcal{L}_{\text{total}} = \mathcal{L}_{\text{CFM}} + \lambda_{\text{align}} \mathcal{L}_{\text{align}}.
\end{equation}
The alignment loss weight is set to $\lambda_{\text{align}}$ = 0.5, balancing the flow matching loss and alignment objective. 

Another challenge in representation alignment from the sequence-length mismatch between the latent representations and the pretrained audio encoder representations. Unlike the image domain, where spatial dimensions can be adjusted through interpolation without disrupting structural integrity, audio representations such as mel-spectrograms and waveform embeddings cannot be resized or interpolated without distorting essential spectral and temporal patterns. To address this, we explore three strategies:

\begin{itemize}
    \item \textbf{Global Pooling:} Applies average pooling to both the latent representation and encoder output to produce one averaged representation. This simplifies alignment but may lose temporal granularity.
    \item \textbf{Interpolation-based Matching:} Interpolates the latent representations to match the encoder's sequence length, preserving temporal structure and alignment.
    \item \textbf{Convolution-based Projection:} We introduce a projection mechanism using a 1D convolutional layer with a kernel size of 1 to reshape the latent representation into the same sequence length as the encoder embedding while preserving temporal dependencies:
    \begin{equation}
    h'_{\phi}(z_t) = \text{Conv1D} (h_{\phi}(z_t)),
    \end{equation}
    where the 1D convolution adjusts \( N_z \) to match \( N_a \), effectively bridges sequence-length differences, enabling precise feature alignment across timesteps.
\end{itemize}

By modulating alignment dynamically and resolving sequence-length mismatches, TRA ensures that latent representations evolve meaningfully throughout the synthesis process, enhancing both perceptual quality and temporal synchronization in video-to-audio generation.

\begin{table*}
    \begin{center}
    \setlength\tabcolsep{6pt} 
    \renewcommand{\arraystretch}{1.1}
    \scalebox{.77}{
    \begin{tabular}{l | c c c c c c c c c c c}
        \toprule
        {Methods} & FD$\downarrow$ & FAD$\downarrow$ & FID$\downarrow$ & IS$\uparrow$ & KL$\downarrow$ & Acc(\%) $\uparrow$ & CLIP $\uparrow$  & MOS-Q$\dag$ $\uparrow$ &  MOS-A$\dag$ $\uparrow$  & Infer(s)$\downarrow$  & Params(M)$\downarrow$ \\ 
        \bottomrule
        SpecVQGAN (BMVC'21) \cite{iashin2021taming}         & 1.25 & 5.65 & 22.77 & 15.76 & 7.25 & 57.20 & 6.37 & 2.52 $\pm$ 0.20 & 2.64 $\pm$ 0.24 & 3.58 & 308 \\
        Im2Wav (ICASSP'23) \cite{sheffer2023hear}             & 1.99 & 9.37 & 24.30 & 21.19 & 6.45 & 65.56 & 9.04 & 2.24 $\pm$ 0.23 & 2.81 $\pm$ 0.26 & 218.40 & 448  \\
        Diff-Foley (NeurIPS'23) \cite{luo2024diff}             & 3.34 & 4.71 & \ag{10.55} & \au{56.67} & 6.05 & 93.92 & 8.92 & 2.87 $\pm$ 0.29 & \cu{3.36 $\pm$ 0.29} & 3.00 & 860  \\
        See \& Hear (CVPR'24) \cite{xing2024seeing}         & 2.27 & 5.55 & 21.35 & 19.23 & 6.94 & 58.14 & 8.30 & 2.79 $\pm$ 0.23 & 2.84 $\pm$ 0.19 & 18.25 & 1099  \\
        FoleyCrafter (arXiv) \cite{zhang2024foleycrafter} & \ag{0.99} & 2.45 & 12.07 & \cu{42.06} & \ag{5.67} & 83.54  & \cu{13.47} & \cu{3.18 $\pm$ 0.20} & 3.20 $\pm$ 0.20 & 2.96 & 1252  \\
        Frieren (NeurIPS'24) \cite{wang2025frieren}            & 1.22 & \ag{1.33} & \cu{11.72} & 38.56 & \au{5.60} & \ag{97.15} & \ag{13.59} & \ag{3.20 $\pm$ 0.27} & \ag{3.43 $\pm$ 0.24} & \au{0.52} & \cu{372}  \\
        MDSGen (ICLR'25) \cite{pham2024mdsgen}              & \cu{1.04} & \cu{1.34} & 11.83 & 30.37 & 6.03 & \cu{96.52} & 11.41 & 3.05 $\pm$ 0.29 & 3.21 $\pm$ 0.24 & \cu{1.56} & \au{142}  \\
        \midrule
        \textbf{TARO (Ours)} & \au{0.47} & \au{0.94} & \au{8.21} & \ag{56.60} & \cu{5.71} & \au{97.19} & \au{14.10} & \au{3.87 $\pm$ 0.24} & \au{4.16 $\pm$ 0.21} & \ag{0.58} & \ag{258}  \\
        \bottomrule 
    \end{tabular}
    }
    \vspace{-0.2cm}
    \caption{\textbf{Quantitative comparisons of video-to-audio models on the VGGSound dataset.} We use colors to denote the \au{first}, \ag{second} and \cu{third} places respectively. $\dag$: MOS-AQ and MOS-AV scores are evaluated on samples from both VGGSound and Landscape datasets.}
    \label{table:vggsound_quantitative}
    \vspace{-0.4cm}
    \end{center}
\end{table*}

\begin{figure*}
    \centering
    \includegraphics[width=\linewidth,keepaspectratio]{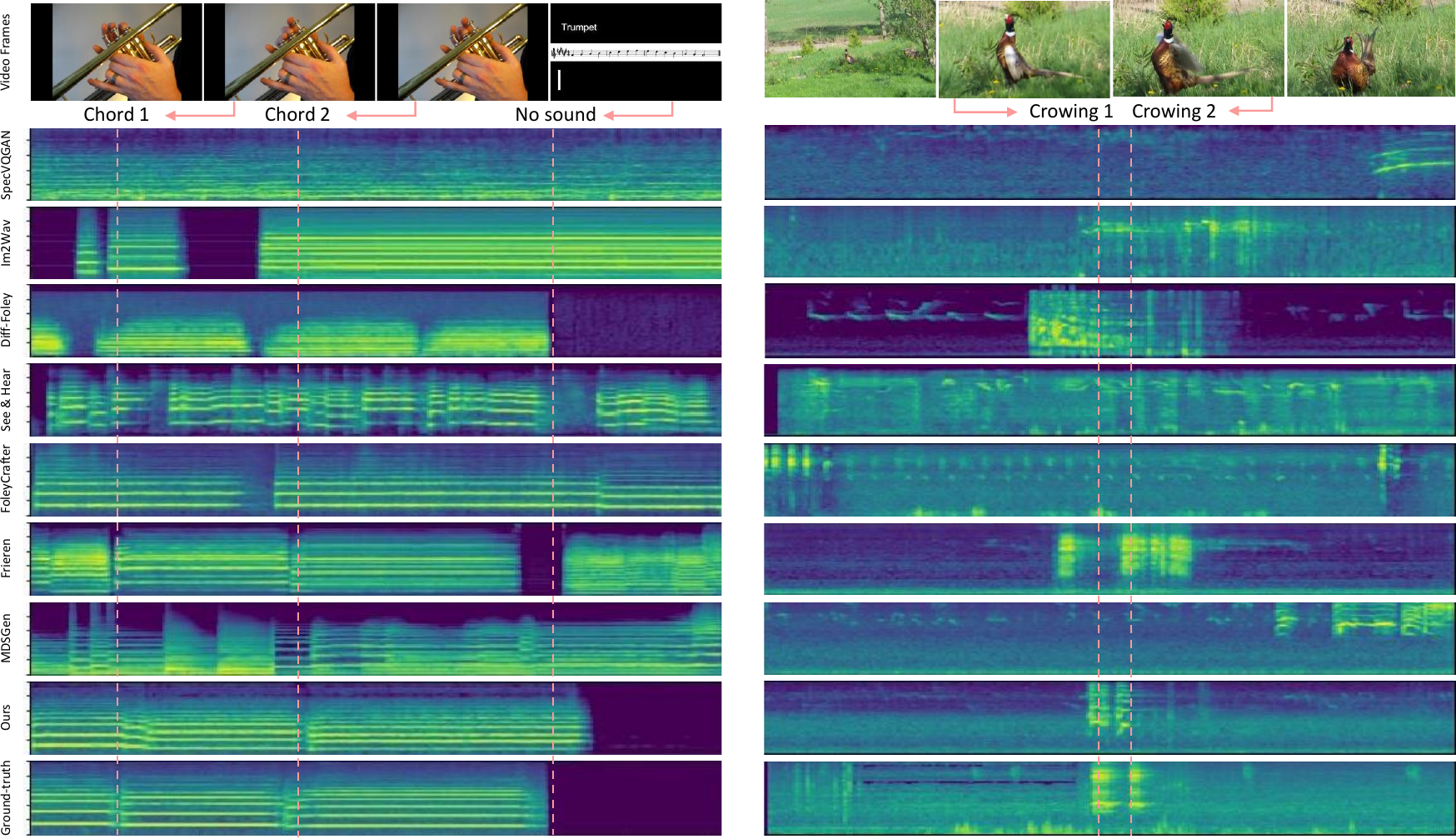}
    \caption{\textbf{Quantitative comparisons of video-to-audio models.} Our TARO achieves superior synchronization and fidelity, closely aligning with the ground truth, while other methods often produce misaligned or extraneous audio, underscoring its effectiveness in capturing event-driven acoustic details.}
    \label{fig:quantitative}
    \vspace{-0.4cm}
\end{figure*}

\section{Experiments}
\label{sec:exp}

\subsection{Experiment Setup}

\noindent \textbf{Datasets} We train our model on the VGGSound \cite{chen2020vggsound} dataset, which comprises 174,337 videos, each lasting 10 seconds, following the official training split. We truncate all videos to 8 seconds and downsample them to 4 FPS, as done in prior work \cite{luo2024diff}. The VGGSound \cite{chen2020vggsound} dataset is a diverse dataset encompassing a broad range of audio-visual events, making it well-suited for training models that generalize across different acoustic environments. For evaluation, we use the original test split of VGGSound, containing 14,336 videos, which we also truncate to 8 seconds to match the training setup.
To assess the generalization capability of our model beyond the training domain, we conduct a zero-shot evaluation on the Landscape dataset \cite{lee2022sound}, which features 1,000 high-fidelity 10-second videos depicting natural environmental scenes. 
To maintain consistency, we truncate all videos to 8 seconds for evaluation.

\noindent \textbf{Model Configuration}
Our Timestep-Adaptive Representation Alignment with Onset-Aware Conditioning (TARO) is a flow-based multimodal transformer with 12 transformer blocks and a hidden dimension of 768. We train the model for 500,000 steps on an NVIDIA A100 GPU using a batch size of 64 and a learning rate of 1e-4, optimized with AdamW \cite{loshchilov2017decoupled}. 
Only the Denoising Flow-based Transformer and Convolutional Projection are trained; other modules, such as the onset detection model, visual encoder, pretrained audio encoder, and VAE encoder/decoder, are frozen.
To enable classifier-free guidance (CFG) \cite{ho2022classifier}, we randomly replace conditioning inputs with a zero vector 10\% of the time during training, improving robustness and generation quality. For audio processing, we leverage a frozen pretrained Variational Autoencoder (VAE) from AudioLDM2 \cite{liu2024audioldm} to encode and decode mel-spectrograms, paired with its vocoder for waveform generation.
In our TRA strategy, we inject pretrained audio encoder knowledge into the 4th transformer blocks, employing EAT \cite{chen2024eat} as the audio encoder. We use a convolution-based projection to address sequence-length mismatches.
For onset detection, we utilize a pretrained model from \cite{comunita2024syncfusion} to extract onset cues, processing 2-second segments from videos downsampled to 15 FPS. During inference, we employ the SDE Euler-Maruyama sampler with 25 sampling steps and a CFG scale of 8.0 for optimal generation quality.

\noindent \textbf{Metrics}
We evaluate our audio generation model using a comprehensive set of metrics.
For objective evaluations, we compute Frechet Distance (FD) and Frechet Audio Distance (FAD) from AudioLDM \cite{liu2023audioldm}, alongside Frechet Inception Distance (FID), Inception Score (IS), and Mean KL Divergence (KL) using SpecVQGAN \cite{iashin2021taming} scripts. Alignment Accuracy, based on scripts from prior works \cite{luo2024diff}, quantifies audio-visual synchronization. Additionally, we measure perceptual similarity between video frames and generated audio using CLIP Score, computed with Wav2CLIP \cite{wu2022wav2clip} for audio and CLIP \cite{radford2021learning} for video. We extract one video frame every 30 frames, average the visual features, and compute similarity with the corresponding audio features. 
The inference time is conducted on a single NVIDIA A100 GPU with a batch size of 1.
For subjective evaluation, we perform a user study where participants rate generated audio samples using 1-5 Likert scales \cite{likert1932technique}. We report Mean Opinion Scores (MOS) for audio quality (MOS-Q) and content alignment (MOS-A), providing 95\% confidence intervals. To ensure diversity in evaluation, we assess samples from VGGSound \cite{chen2020vggsound} and Landscape \cite{lee2022sound}, covering a wide range of audio-visual scenarios. Additional details of evaluation are in \textit{Suppl} Sec. \ref{sec:subjective_eval}.

\begin{table}[t]
    \begin{center}
    \setlength\tabcolsep{6pt} 
    \renewcommand{\arraystretch}{1.1}
    \scalebox{.8}{
    \begin{tabular}{l | c c c c c}
        \toprule
        {Methods} & FD$\downarrow$ & FAD$\downarrow$ & Acc(\%) $\uparrow$ & CLIP $\uparrow$ \\ 
        \bottomrule
        SpecVQGAN (BMVC'21) \cite{iashin2021taming}         & \cu{2.71} & 8.23 & 74.80 & 6.55  \\
        Im2Wav (ICASSP'23) \cite{sheffer2023hear}           & 4.24 & 9.03 & 78.10 & 8.47  \\
        Diff-Foley (NeurIPS'23) \cite{luo2024diff}          & 7.99 & 5.55 & 84.40 & 8.75  \\
        See \& Hear (CVPR'24) \cite{xing2024seeing}         & 5.43 & 8.25 & 51.40 & 5.75  \\
        FoleyCrafter (arXiv) \cite{zhang2024foleycrafter}   & 3.93 & \au{3.82} & 85.20 & 8.45  \\
        Frieren (NeurIPS'24) \cite{wang2025frieren}         & \ag{2.60} & \cu{5.45} & \ag{97.70} & \cu{9.15}  \\
        MDSGen (ICLR'25) \cite{pham2024mdsgen}              & 3.80 & 6.13 & \cu{97.40} & \ag{10.53}  \\
        \midrule
        \textbf{TARO (Ours)}                                & \au{2.53} & \ag{5.41} & \au{98.20} & \au{10.68}  \\
        \bottomrule 
    \end{tabular}
    }
    \vspace{-0.2cm}
    \caption{\textbf{Quantitative comparisons of video-to-audio models on the Landscape dataset.}}
    \label{table:landscape_quantitative}
    \vspace{-0.5cm}
    \end{center}
\end{table}

\noindent \textbf{Baseline Models} We evaluate our TARO against SOTA video-to-audio models, including SpecVQGAN \cite{iashin2021taming}, Im2Wav \cite{sheffer2023hear}, Diff-Foley \cite{luo2024diff}, See \& Hear \cite{xing2024seeing}, FoleyCrafter \cite{zhang2024foleycrafter}, Frieren \cite{wang2025frieren}, and MDSGen \cite{pham2024mdsgen}. For MDSGen, we select the Audio-VAE variant, as it offers significantly superior audio quality while maintaining high alignment accuracy compared to the Image-VAE version. We conduct both quantitative and qualitative evaluations to assess audio fidelity, semantic relevance, and temporal synchronization. The comparison provides a comprehensive analysis of how TARO performs relative to existing approaches across these key dimensions.

\subsection{Main Results}
\label{sec:quanti}

Table \ref{table:vggsound_quantitative} compares our TARO with SOTA video-to-audio methods \cite{iashin2021taming, sheffer2023hear, luo2024diff, xing2024seeing, zhang2024foleycrafter, wang2025frieren, pham2024mdsgen} on the VGGSound dataset \cite{chen2020vggsound}. TARO outperforms all baselines in FD, FAD, and alignment accuracy, delivering high-fidelity, semantically coherent, and well-synchronized audio. It also achieves the highest CLIP Score, indicating strong perceptual alignment with video content.
Diffusion-based models like Diff-Foley \cite{luo2024diff} and FoleyCrafter \cite{zhang2024foleycrafter} produce competitive results but suffer from high computational costs. Flow-based approaches such as Frieren \cite{wang2025frieren} and masked diffusion models like MDSGen \cite{pham2024mdsgen} improve efficiency but trade off quality and synchronization. In contrast, TARO balances performance across all metrics while maintaining fast inference (0.58s) and a compact model size (258M parameters), making it practical for real-time applications.
Human evaluations further validate TARO’s effectiveness, achieving the highest Mean Opinion Scores with MOS-Q 3.87 and MOS-A 4.16, confirming its ability to generate perceptually natural, well-synchronized audio.

\begin{figure}
    \centering
    \includegraphics[width=\linewidth,keepaspectratio]{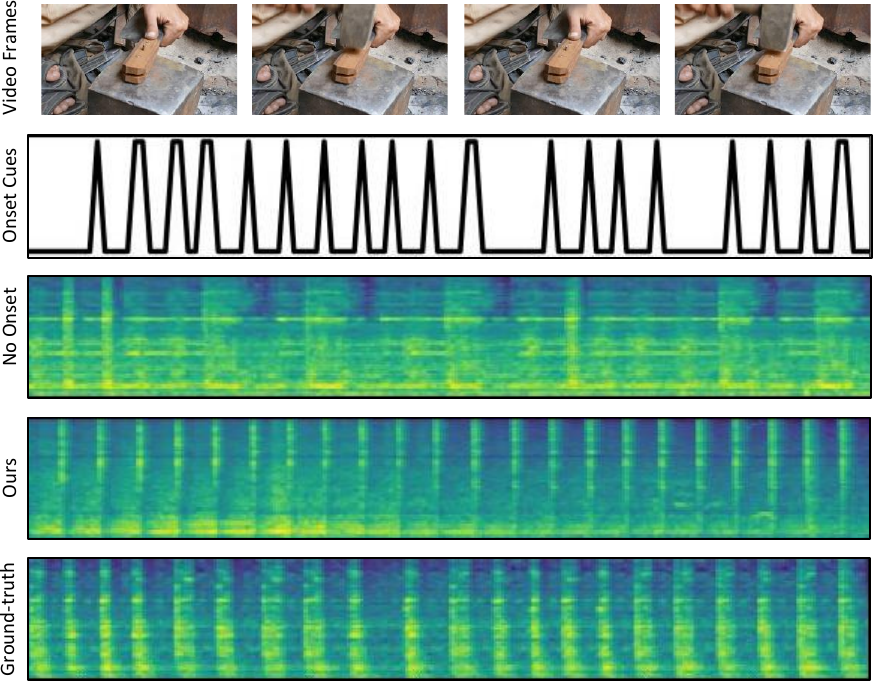}
    \caption{\textbf{Impact of Onset-Aware Conditioning on Audio Synchronization.} This figure compares generated spectrograms with and without OAC.}
    \label{fig:ablation}
\end{figure}

\begin{table}[t]
    \centering
    \vspace{-0.3cm}
    \scalebox{1.}{
    \begin{tabular}{c c | c c c c}
        \toprule
        OAC & TRA & FD$\downarrow$ & FAD$\downarrow$ & Acc(\%) $\uparrow$ & CLIP $\uparrow$ \\ 
        \bottomrule
                   &            & 0.55 & 1.24 & 97.04 & 13.73  \\
                   & \checkmark & \ag{0.51} & \au{0.78} & 97.13 & 13.58  \\
        \checkmark &            & 0.53      & 1.10 & \ag{97.16} & \ag{14.05}  \\
        \checkmark & \checkmark & \au{0.47}   & \ag{0.94} & \au{97.19} & \au{14.10}  \\
        \bottomrule 
    \end{tabular}
    }
    \vspace{-0.2cm}
    \caption{\textbf{Ablation study on key components of TARO.}}
    \label{table:mainablation}
    \vspace{-0.4cm}
\end{table}

Figure \ref{fig:quantitative} presents the qualitative comparison of our TARO and SOTA video-to-audio methods \cite{iashin2021taming, sheffer2023hear, luo2024diff, xing2024seeing, zhang2024foleycrafter, wang2025frieren, pham2024mdsgen}. TARO exhibits superior audio quality and synchronization, closely aligning generated spectrograms with ground-truth. On the right, it effectively reproduces two distinct crowing events in a pheasant video, ensuring seamless alignment with visual cues and enhancing perceptual coherence. These results highlight TARO’s ability to model fine-grained audio-visual correlations more effectively than prior methods. Additional audio samples and visualizations are provided in \textit{Suppl} Sec. \ref{sec:suppl_visualization}.

Table \ref{table:landscape_quantitative} presents a quantitative comparison of TARO against state-of-the-art video-to-audio methods \cite{iashin2021taming, sheffer2023hear, luo2024diff, xing2024seeing, zhang2024foleycrafter, wang2025frieren, pham2024mdsgen} on the Landscape dataset \cite{lee2022sound}. TARO achieves the highest synchronization accuracy and CLIP score, demonstrating superior alignment between generated audio and visual content. It also maintains competitive audio fidelity, closely matching the best-performing methods. These results highlight TARO’s strong generalization to diverse natural scenes, producing high-quality, well-synchronized audio without fine-tuning.

\begin{table}[t]
    \centering
    \scalebox{.95}{
    \begin{tabular}{l | c c c c}
        \toprule
         & FD$\downarrow$ & FAD$\downarrow$ & Acc(\%) $\uparrow$ & CLIP $\uparrow$ \\ 
        \bottomrule
        w/o weighted           & 0.51 & 0.97 & 97.15 & 14.04  \\
        w/ weighted \textbf{(Ours)}         & \au{0.47} & \au{0.94} & \au{97.19} & \au{14.10}  \\
        \midrule
        Global Pooling        & 0.51 & 1.16 & 97.06 & 14.01 \\
        Interpolate           & \ag{0.49} & \ag{1.09} & \ag{97.13} & \au{14.12}  \\
        Conv \textbf{(Ours)}         & \au{0.47} & \au{0.94} & \au{97.19} & \ag{14.10}  \\
        \bottomrule 
    \end{tabular}
    }
    \vspace{-0.2cm}
    \caption{\textbf{Ablation study on the details of Timestep-Adaptive Representation Alignment.}}
    \label{table:ablationalign}
    \vspace{-0.4cm}
\end{table}

\subsection{Ablation Studies}

We conduct an ablation study on the VGGSound \cite{chen2020vggsound} test set to assess the impact of OAC and TRA in our TARO. Our full model integrates both components using a convolution-based projection with the EAT \cite{chen2024eat} audio encoder at the 4th transformer block.

As shown in Table \ref{table:mainablation}, removing both components leads to the lowest scores. TRA alone improves distribution matching (FD, FAD), while OAC enhances synchronization (Acc, CLIP Score). Combining both achieves the best results, demonstrating their complementary effects on audio quality and alignment.
Figure \ref{fig:ablation} visualizes the impact of OAC. Without it, the generated spectrogram lacks a clear temporal structure and deviates from the ground truth. In contrast, incorporating OAC, our model closely aligns with the ground truth, ensuring precise synchronization between generated audio and visual events.
Full metric evaluations for all ablation are provided in \textit{Suppl} Sec. \ref{sec:suppl_ablation}.

\noindent \textbf{Ablation on Timestep-Adaptive Representation Alignment} To assess the impact of TRA, we compare weighting strategies and different sequence alignment methods in Table \ref{table:ablationalign}. Removing adaptive weighting reduces performance across all metrics, confirming its role in dynamically adjusting alignment strength. Global pooling weakens temporal granularity for sequence-length alignment, while interpolation improves CLIP Score but falls short in distribution matching. Our convolution-based projection achieves the best balance, preserving temporal dependencies and ensuring robust alignment. These results highlight the effectiveness of adaptive weighting and convolution-based alignment in enhancing audio fidelity and synchronization.

\noindent \textbf{Ablation on Different Audio Feature Extractor} We examine the effect of different pretrained audio encoders on TRA. Table \ref{table:ablationencoder} presents a comparison among wav2vec 2.0 \cite{baevski2020wav2vec}, CLAP \cite{wu2023large}, BEATs \cite{chen2022beats}, and EAT \cite{chen2024eat}. While all encoders contribute to strong performance, EAT consistently achieves the best results across FD, alignment accuracy, and CLIP Score while maintaining the lowest FAD. These results highlight EAT’s effectiveness in providing high-quality audio representations, making it the optimal choice for TARO by enhancing synchronization and fidelity in video-to-audio synthesis.

\begin{table}[t]
    \centering
    \vspace{-0.3cm}
    \scalebox{1}{
    \begin{tabular}{l | c c c c}
        \toprule
        {Encoder} & FD$\downarrow$ & FAD$\downarrow$ & Acc(\%) $\uparrow$ & CLIP $\uparrow$ \\ 
        \bottomrule
        wav2vec 2.0 \cite{baevski2020wav2vec}  & \ag{0.48} & 1.41 & \ag{97.15} & \ag{14.04} \\
        CLAP \cite{wu2023large}                & 0.49 & \ag{0.81} & 96.58 & 13.85  \\
        BEATs \cite{chen2022beats}             & 0.52 & \au{0.74} & 97.13 & 13.41  \\
        EAT \cite{chen2024eat} \textbf{(Ours)} & \au{0.47} & 0.94 & \au{97.19} & \au{14.10}  \\
        \bottomrule 
    \end{tabular}
    }
    \vspace{-0.2cm}
    \caption{\textbf{Ablation study on different audio encoders.}}
    \label{table:ablationencoder}
\end{table}

\begin{table}[t]
    \centering
    \vspace{-0.3cm}
    \scalebox{1.}{
    \begin{tabular}{l | c c c c}
        \toprule
        {Depth} & FD$\downarrow$ & FAD$\downarrow$ & Acc(\%) $\uparrow$ & CLIP $\uparrow$ \\ 
        \bottomrule
        2                 & 0.49 & 1.49 & \ag{97.13} & \au{14.14}  \\
        4 \textbf{(Ours)} & \au{0.47} & 0.94 & \au{97.19} & \ag{14.10}  \\
        6                 & \ag{0.48} & 0.82 & 97.11 & 13.75  \\
        8                 & 0.48 & \au{0.77} & 96.92 & 13.73  \\
        10                & 0.49 & \ag{0.79} & 97.09 & 13.63   \\
        \bottomrule 
    \end{tabular}
    }
    \vspace{-0.2cm}
    \caption{\textbf{Ablation study on the effect of injecting audio encoder features at different transformer blocks.}}
    \label{table:ablationlayer}
    \vspace{-0.4cm}
\end{table}

\noindent \textbf{Ablation on Audio Encoder Injection Depth}
We analyze the effect of injecting audio encoder knowledge at different transformer blocks (2nd, 4th, 6th, 8th, and 10th) within TRA. As shown in Table \ref{table:ablationlayer}, the 4th block achieves the best FD and alignment accuracy while maintaining a strong CLIP Score and competitive FAD. Injecting too early (2nd) or too late (6th, 8th, 10th) degrades performance, indicating that the 4th block offers the optimal balance between early integration and sufficient feature refinement, ensuring effective audio-visual synchronization without overfitting or underutilization of encoder features.

\section{Conclusion}
We present TARO, a novel Timestep-Adaptive Representation Alignment with Onset-Aware Conditioning framework for video-to-audio synthesis. By integrating Timestep-Adaptive Representation Alignment (TRA) and Onset-Aware Conditioning (OAC), TARO effectively aligns audio representations with pretrained audio encoders while capturing event-driven cues for precise synchronization.
Our framework efficiently generates high-fidelity, temporally aligned audio from silent videos while maintaining computational efficiency. Benchmark evaluations demonstrate that TARO outperforms SOTA methods in fidelity, synchronization, and perceptual quality.
\section*{Acknowledgments}
This work was partly supported by Institute of Information \& communications Technology Planning \& Evaluation (IITP) grant funded by the Korea government(MSIT) [RS-2021-II212068, Artificial Intelligence Innovation Hub (Seoul National University)] and Artificial intelligence industrial convergence cluster development project funded by the Ministry of Science and ICT(MSIT, Korea) \& Gwangju Metropolitan City.
{
    \small
    \bibliographystyle{ieeenat_fullname}
    \bibliography{main}
}

\clearpage
\setcounter{page}{1}
\maketitlesupplementary

\appendix

\section{Demo Audios}
We recommend that readers refer to our project page at {\href{https://github.com/triton99/TARO}{\nolinkurl{github.com/triton99/TARO}}},
showcasing extensive qualitative comparisons between our TARO and SOTA video-to-audio generation methods \cite{iashin2021taming, sheffer2023hear, luo2024diff, xing2024seeing, zhang2024foleycrafter, wang2025frieren, pham2024mdsgen}. Please note that since the provided project page for this supplementary material is \textit{offline}, and therefore, \textit{no modifications can be made after submission}; it is offered solely for the convenience of visualization. The project page features various demo video-to-audio synthesis, including comparisons for our TARO and prior methods \cite{iashin2021taming, sheffer2023hear, luo2024diff, xing2024seeing, zhang2024foleycrafter, wang2025frieren, pham2024mdsgen} on both VGGSound \cite{chen2020vggsound} and Landscape \cite{lee2022sound} datasets. 

\begin{table}[t]
    \centering
    \vspace{-0.3cm}
    \scalebox{.7}{
    \begin{tabular}{c c | c c c c c c c}
        \toprule
        OAC & TRA & FD$\downarrow$ & FAD$\downarrow$ & FID$\downarrow$ & IS$\uparrow$ & KL$\downarrow$ & Acc(\%) $\uparrow$ & CLIP $\uparrow$ \\ 
        \bottomrule
                   &            & 0.55 & 1.24 & 8.80 & 50.01 & 5.97 & 97.04 & 13.73  \\
                   & \checkmark & \ag{0.51} & \au{0.78} & \ag{8.65} & 52.43 & \ag{5.86} & 97.13 & 13.58  \\
        \checkmark &            & 0.53      & 1.10 & 8.85 & \ag{52.65} & 5.98 & \ag{97.16} & \ag{14.05}  \\
        \checkmark & \checkmark & \au{0.47}   & \ag{0.94} & \au{8.21} & \au{56.60} & \au{5.71}  & \au{97.19} & \au{14.10}  \\
        \bottomrule 
    \end{tabular}
    }
    \vspace{-0.2cm}
    \caption{\textbf{Ablation study on key components of TARO.}}
    \label{table:mainablation_full}
    \vspace{-0.4cm}
\end{table}

\begin{table}[t]
    \centering
    \scalebox{.65}{
    \begin{tabular}{l | c c c c c c c}
        \toprule
         & FD$\downarrow$ & FAD$\downarrow$ & FID$\downarrow$ & IS$\uparrow$ & KL$\downarrow$ & Acc(\%) $\uparrow$ & CLIP $\uparrow$ \\ 
        \bottomrule
        w/o weighted           & 0.51 & 0.97 & 8.34 & 53.36 & 5.84 & 97.15 & 14.04  \\
        w/ weighted \textbf{(Ours)}         & \au{0.47} & \au{0.94} & \au{8.21} & \au{56.60} & \au{5.71} & \au{97.19} & \au{14.10}  \\
        \midrule
        Global Pooling        & 0.51 & 1.16 & 9.21 & 50.53 & 5.94 & 97.06 & 14.01 \\
        Interpolate           & \ag{0.49} & \ag{1.09} & \ag{8.23} & \au{57.01} & \ag{5.73} & \ag{97.13} & \au{14.12}  \\
        Conv \textbf{(Ours)}         & \au{0.47} & \au{0.94} & \au{8.21} & \ag{56.60} & \au{5.71} & \au{97.19} & \ag{14.10}  \\
        \bottomrule 
    \end{tabular}
    }
    \vspace{-0.2cm}
    \caption{\textbf{Ablation study on the details of Timestep-Adaptive Representation Alignment.}}
    \label{table:ablationalign_full}
    \vspace{-0.4cm}
\end{table}

\section{Subjective evaluation}
\label{sec:subjective_eval}

To comprehensively evaluate the performance of our Timestep-Adaptive Representation Alignment with Onset-Aware Conditioning (TARO) in video-to-audio synthesis, we conduct a user study involving 20 videos from the VGGSound dataset \cite{chen2020vggsound} and 10 videos from the Landscape dataset \cite{lee2022sound}. Our study focuses on two key aspects: audio quality and temporal alignment.
For audio quality assessment (Mean Opinion Score for Audio Quality, MOS-Q), participants are instructed to focus exclusively on the generated audio, disregarding the accompanying video content, and rate its perceptual quality. This ensures that judgments reflect the naturalness and clarity of the audio without being influenced by visual cues.
For temporal alignment evaluation (Mean Opinion Score for Content Alignment, MOS-A), participants assess how well the generated audio synchronizes with the corresponding video, ignoring audio quality to provide an unbiased measure of synchronization accuracy.
To mitigate potential biases, we group samples by video and randomly shuffle their order within each group before presenting them to participants. Each sample is rated on a 1–5 Likert scale \cite{likert1932technique}, offering a robust subjective evaluation of TARO’s capability to generate high-quality and well-aligned audio in video-to-audio synthesis. Our study involved 20 participants, both male and female, aged from 24 to 30, primarily graduate students, with 50\% having experience in generative models.

\section{Comprehensive Metric Analysis for Ablation Study}
\label{sec:suppl_ablation}

In addition to the primary metrics discussed in the main paper, Table \ref{table:mainablation_full} presents a detailed comparison of FID, IS, and KL scores across different ablation settings. The results show that Timestep-Adaptive Representation Alignment (TRA) improves distributional matching, reducing FID and KL while increasing IS, indicating enhanced generative quality. Onset-Aware Conditioning (OAC) primarily benefits perceptual alignment, contributing to a higher IS score. The full TARO, integrating both components, achieves the best performance across all metrics, highlighting their complementary role in improving fidelity, synchronization, and generative quality in video-to-audio synthesis.

\begin{table}[t]
    \centering
    \vspace{-0.3cm}
    \scalebox{.7}{
    \begin{tabular}{l | c c c c c c c}
        \toprule
        {Encoder} & FD$\downarrow$ & FAD$\downarrow$ & FID$\downarrow$ & IS$\uparrow$ & KL$\downarrow$ & Acc(\%) $\uparrow$ & CLIP $\uparrow$ \\ 
        \bottomrule
        wav2vec 2.0 \cite{baevski2020wav2vec}  & \ag{0.48} & 1.41 & 8.90 & \ag{56.36} & 5.88 & \ag{97.15} & \ag{14.04} \\
        CLAP \cite{wu2023large}                & 0.49 & \ag{0.81} & 8.44 & 52.02 & \au{5.53} & 96.58 & 13.85  \\
        BEATs \cite{chen2022beats}             & 0.52 & \au{0.74} & \ag{8.25} & 53.31 & \ag{5.64} & 97.13 & 13.41  \\
        EAT \cite{chen2024eat} \textbf{(Ours)} & \au{0.47} & 0.94 & \au{8.21} & \au{56.60} & 5.71 & \au{97.19} & \au{14.10}  \\
        \bottomrule 
    \end{tabular}
    }
    \vspace{-0.2cm}
    \caption{\textbf{Ablation study on different audio encoders.}}
    \label{table:ablationencoder_full}
\end{table}

\begin{table}[t]
    \centering
    \vspace{-0.3cm}
    \scalebox{.75}{
    \begin{tabular}{l | c c c c c c c}
        \toprule
        {Depth} & FD$\downarrow$ & FAD$\downarrow$ & FID$\downarrow$ & IS$\uparrow$ & KL$\downarrow$ & Acc(\%) $\uparrow$ & CLIP $\uparrow$ \\ 
        \bottomrule
        2                 & 0.49 & 1.49 & 8.52 & \ag{56.57} & 5.73 & \ag{97.13} & \au{14.14}  \\
        4 \textbf{(Ours)} & \au{0.47} & 0.94 & \au{8.21} & \au{56.60} & \au{5.71} & \au{97.19} & \ag{14.10}  \\
        6                 & \ag{0.48} & 0.82 & 8.71 & 53.02 & 5.83 & 97.11 & 13.75  \\
        8                 & 0.48 & \au{0.77} & \ag{8.51} & 54.39 & \ag{5.72} & 96.92 & 13.73  \\
        10                 & 0.49 & \ag{0.79} & 8.59 & 54.26 & 5.79 & 97.09 & 13.63  \\
        \bottomrule 
    \end{tabular}
    }
    \vspace{-0.2cm}
    \caption{\textbf{Ablation study on the effect of injecting audio encoder features at different transformer blocks.}}
    \label{table:ablationlayer_full}
    \vspace{-0.4cm}
\end{table}

\noindent \textbf{Ablation on Timestep-Adaptive Representation Alignment}
As shown in Table \ref{table:ablationalign_full}, FID improves with adaptive weighting and convolution-based projection, achieving the lowest value and indicating better distribution alignment with real audio. IS is maximized when interpolation is used, suggesting enhanced diversity, but it comes at the cost of weaker distribution matching. Our convolution-based projection achieves a strong balance, maintaining a high IS score while minimizing KL divergence, demonstrating improved synthesis quality and robustness in audio generation.

\noindent \textbf{Ablation on Different Audio Feature Extractor}
Table \ref{table:ablationencoder_full} shows that EAT \cite{chen2024eat} achieves the lowest FID, indicating superior distribution alignment with real audio compared to other encoders. IS scores vary across encoders, with EAT \cite{chen2024eat} and wav2vec 2.0 \cite{baevski2020wav2vec} performing best, suggesting they enable more diverse and expressive audio generation. KL divergence is lowest with CLAP \cite{wu2023large}, which may indicate smoother latent representations, though its lower accuracy and CLIP Score suggest weaker synchronization. EAT \cite{chen2024eat} maintains a strong balance, achieving competitive KL while excelling in FID and IS, reinforcing its effectiveness in improving fidelity and synchronization.

\noindent \textbf{Ablation on Audio Encoder Injection Depth}
Table \ref{table:ablationlayer_full} demonstrates that injecting at the 4th block yields the lowest FID, confirming its effectiveness in aligning the learned distribution with real audio. While IS remains relatively stable, early injection (2nd block) attains the highest score but also results in a higher FID, indicating potential misalignment. KL divergence is lowest at the 4th block, suggesting it provides the best trade-off between audio-visual integration and representation refinement. Later injections (6th, 8th, and 10th blocks) slightly increase FID and KL, implying reduced effectiveness in synchronizing latent audio features with the video context. 

\section{Additional Results \& Analysis}
\label{sec:suppl_analysis}

\begin{figure}
    \centering
    \includegraphics[width=.85\linewidth,keepaspectratio]{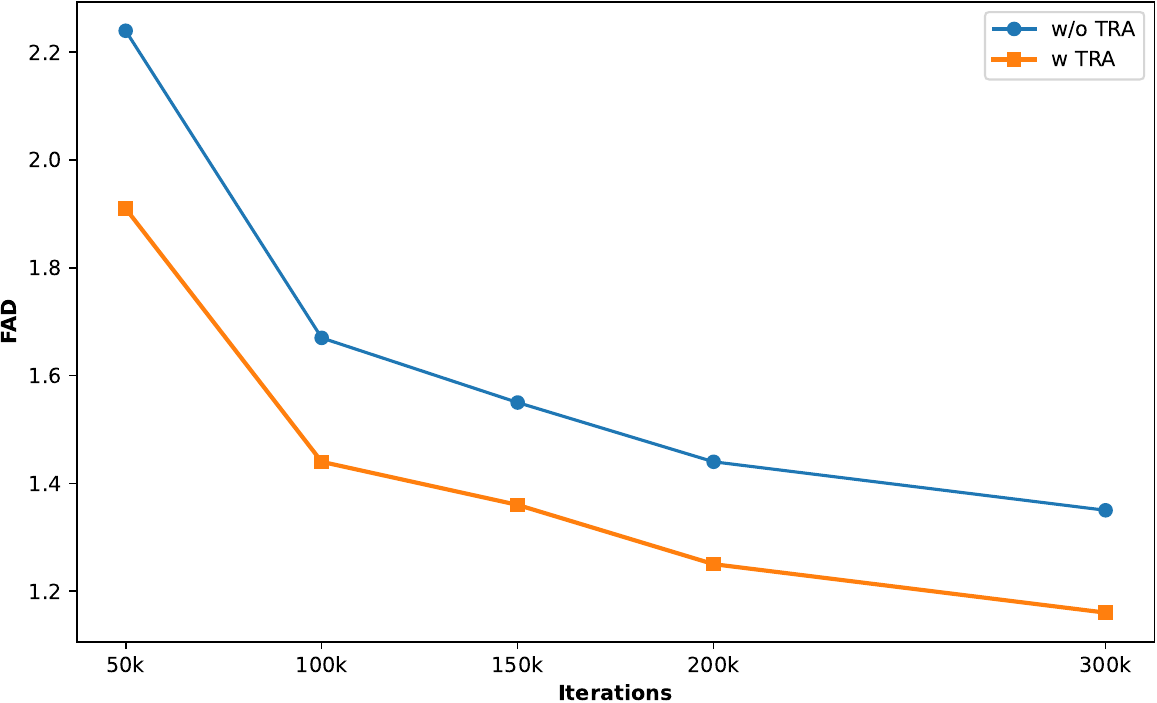}
    \caption{\textbf{Training efficiency.}}
    \label{fig:fad_plot}
\end{figure}

\noindent \textbf{Training efficiency} As shown in Figure \ref{fig:fad_plot}, TRA improves both training stability and generation quality compared to the vanilla model.

\begin{table}[t]
    \centering
    \vspace{-0.3cm}
    \scalebox{.9}{
    \begin{tabular}{l | c c c c}
        \toprule
        {Depth} & FD$\downarrow$ & FAD$\downarrow$ & Acc(\%) $\uparrow$ & CLIP $\uparrow$ \\ 
        \bottomrule
        Energy control & 0.56 & \au{0.89} & 96.10 & 13.92 \\
        Down-sampled mel & 0.59 & 1.02 & 96.90 & 13.88 \\
        Onset cues (Ours) &\au{0.47} & 0.94 &\au{97.19} &\au{14.10} \\
        \bottomrule 
    \end{tabular}
    }
    \vspace{-0.2cm}
    \caption{\textbf{Ablation study on temporal features.}}
    \label{table:ablationtemp}
    \vspace{-0.4cm}
\end{table}

\noindent \textbf{Comparison with other temporal features}. Tab. \ref{table:ablationtemp} shows the effectiveness of different temporal features: energy control \cite{jeong2025read}, down-sampled melspectrogram \cite{wang2024tiva}, and onset cues (Ours). We train all methods with 500k iterations. As shown, onset cues achieve the best balance, yielding the highest synchronization accuracy and CLIP score, with competitive FAD and FD.

\noindent \textbf{Other metric report}. In addition to commonly reported metrics, we provide some recently used metrics such as ImageBind Score, Onset Acc, Onset AP, and Temporal Offset in Tab. \ref{table:ablationmetrics} on the VGGSound dataset.

\section{Limitation}
\label{sec:suppl_limit}

\begin{table}[t]
    \centering
    \vspace{-0.3cm}
    \scalebox{.8}{
    \begin{tabular}{l|c|c|c|c}
    \toprule
    {Method} & IB$\uparrow$ & Onset Acc$\uparrow$ & Onset AP$\uparrow$ & Offsets(s)$\downarrow$ \\
    \midrule
    SpecVQGAN    & 56.43 & 28.09  & 54.70 & 1.18 \\
    Im2Wav       & 55.75 & 27.76  & 52.57 & 1.16 \\
    Diff-Foley   & 59.64 & 25.47  & 61.90 & 1.09 \\
    See \& Hear  & 62.76 & 26.41  & 60.72 & 1.20 \\
    FoleyCrafter & 63.66 & 29.18  & 55.19 & 1.14 \\
    Frieren      & 61.26 & 26.53  & \au{63.72} & 0.98 \\
    MDSGen       & 60.78 & 27.90  & 58.14 & 1.16 \\
    Ours         & \au{64.22} & \au{29.37}  & 62.17 & \au{0.97} \\
    \bottomrule
    \end{tabular}
    }
    \vspace{-0.2cm}
    \caption{\textbf{Other metrics on VGGSound dataset.}}
    \label{table:ablationmetrics}
    \vspace{-0.4cm}
\end{table}

Despite the strong performance of TARO, several limitations remain.  
First, while the VGGSound \cite{chen2020vggsound} dataset provides a diverse set of audio-visual samples, its scale may not fully exploit the potential of our approach. Expanding to larger and more varied datasets could enhance generalization. 
Second, our method is currently constrained to a fixed video length, limiting its adaptability to variable-length sequences or practical applications. 
Third, like existing video-to-audio synthesis models, TARO is primarily designed to generate Foley sounds and struggles with handling human speech, requiring more fine-grained linguistic and phonetic understanding. 
Additionally, the effectiveness of the Onset-Aware Conditioning (OAC) module depends on the accuracy of onset detection models. Errors in onset prediction could lead to misalignment in synthesized audio. 
Addressing these limitations presents exciting directions for future research, including improving adaptability and generalization and expanding beyond Foley sound generation.

\section{Additional Visualizations}
\label{sec:suppl_visualization}

\begin{figure*}
    \centering
    \includegraphics[width=\linewidth,keepaspectratio]{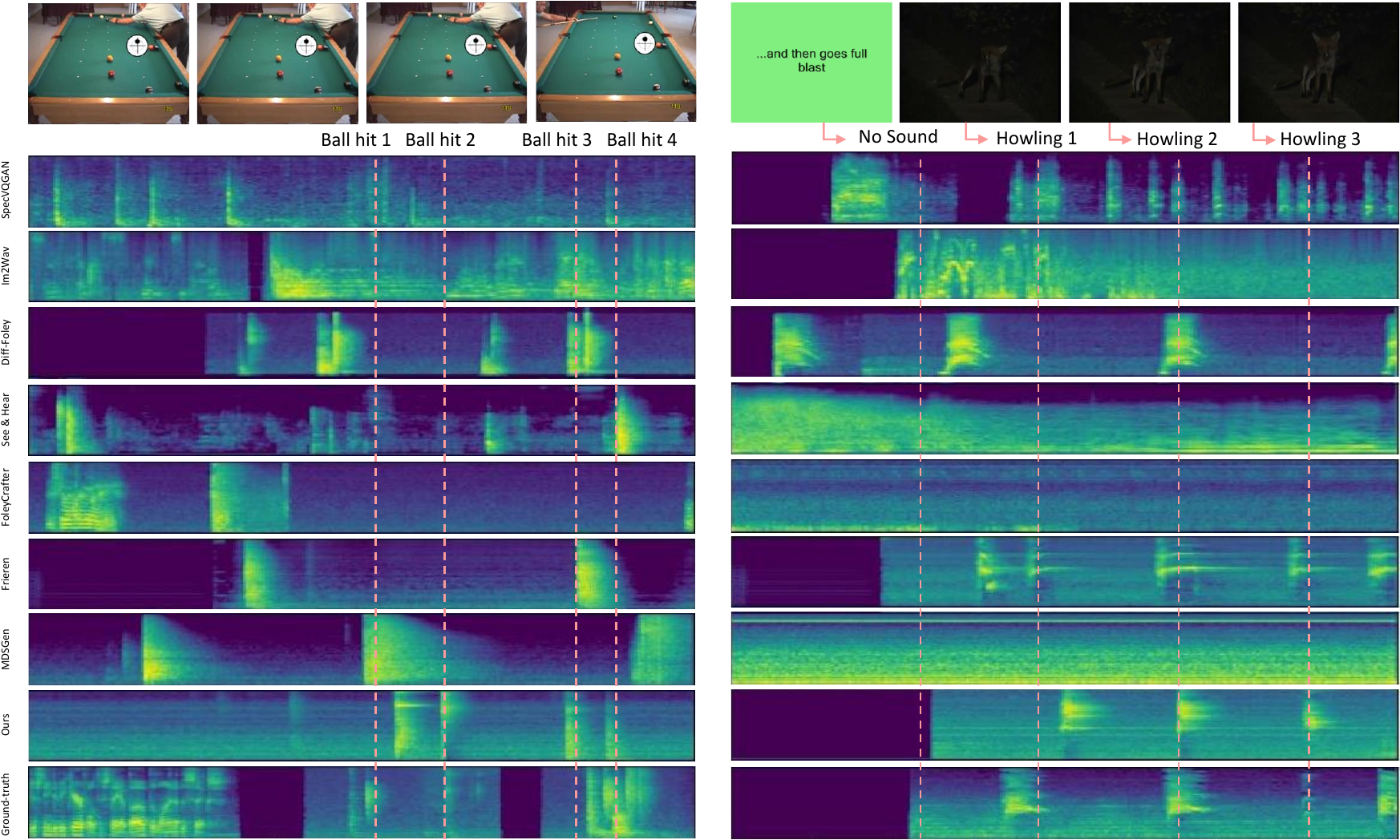}
    \caption{\textbf{Visual comparisons on \textit{1mMqLP36sCQ\_000245} and \textit{1JsIcP2nXMw\_000108} from the VGGSound dataset.}}
    \label{fig:sup_1}
\end{figure*}

\begin{figure*}
    \centering
    \includegraphics[width=\linewidth,keepaspectratio]{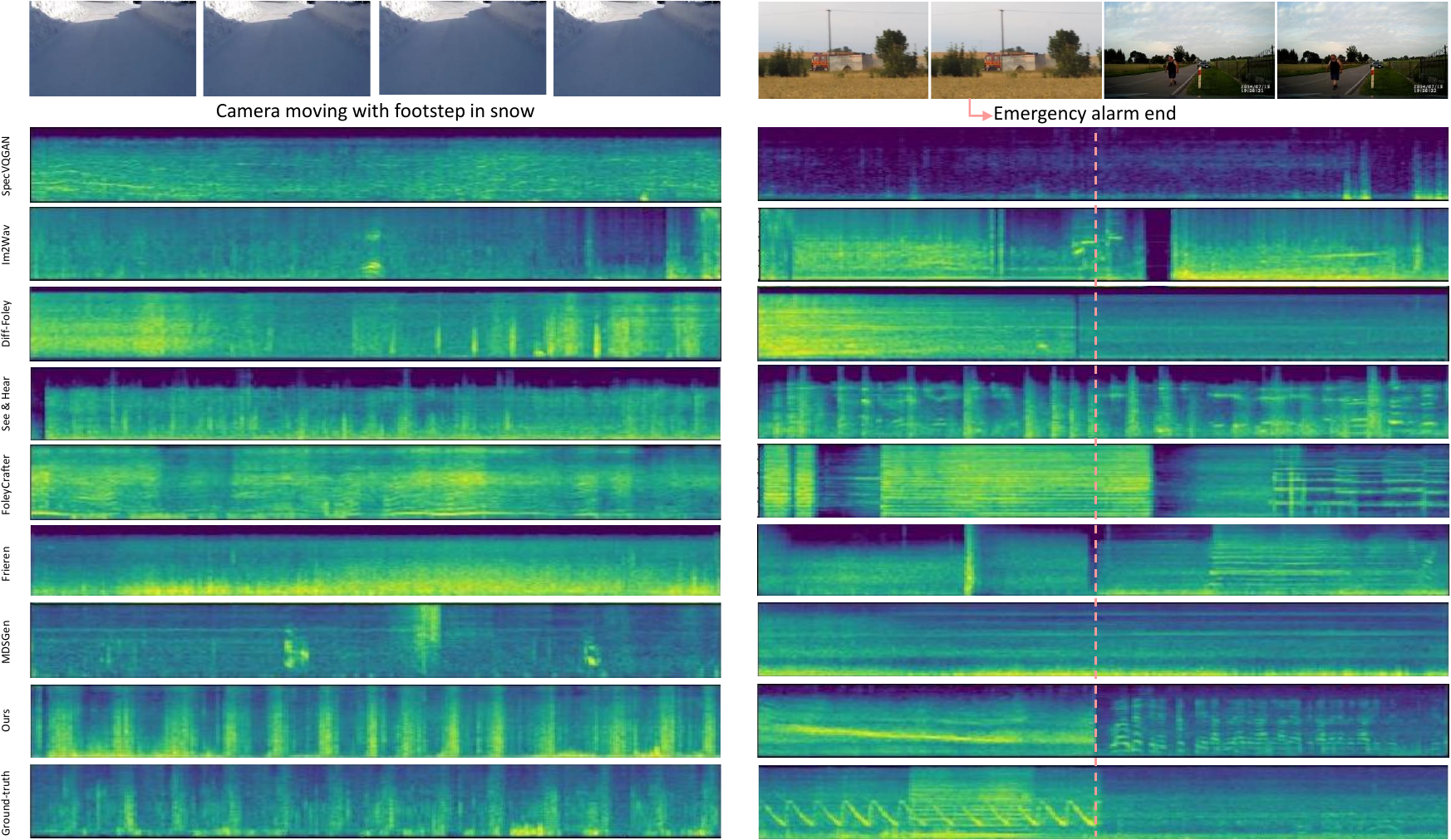}
    \caption{\textbf{Visual comparisons on \textit{KydSULgAHFI\_000084} and \textit{0N6S5OoG7Vg\_000150} from the VGGSound dataset.}}
    \label{fig:sup_2}
\end{figure*}

\begin{figure*}
    \centering
    \includegraphics[width=\linewidth,keepaspectratio]{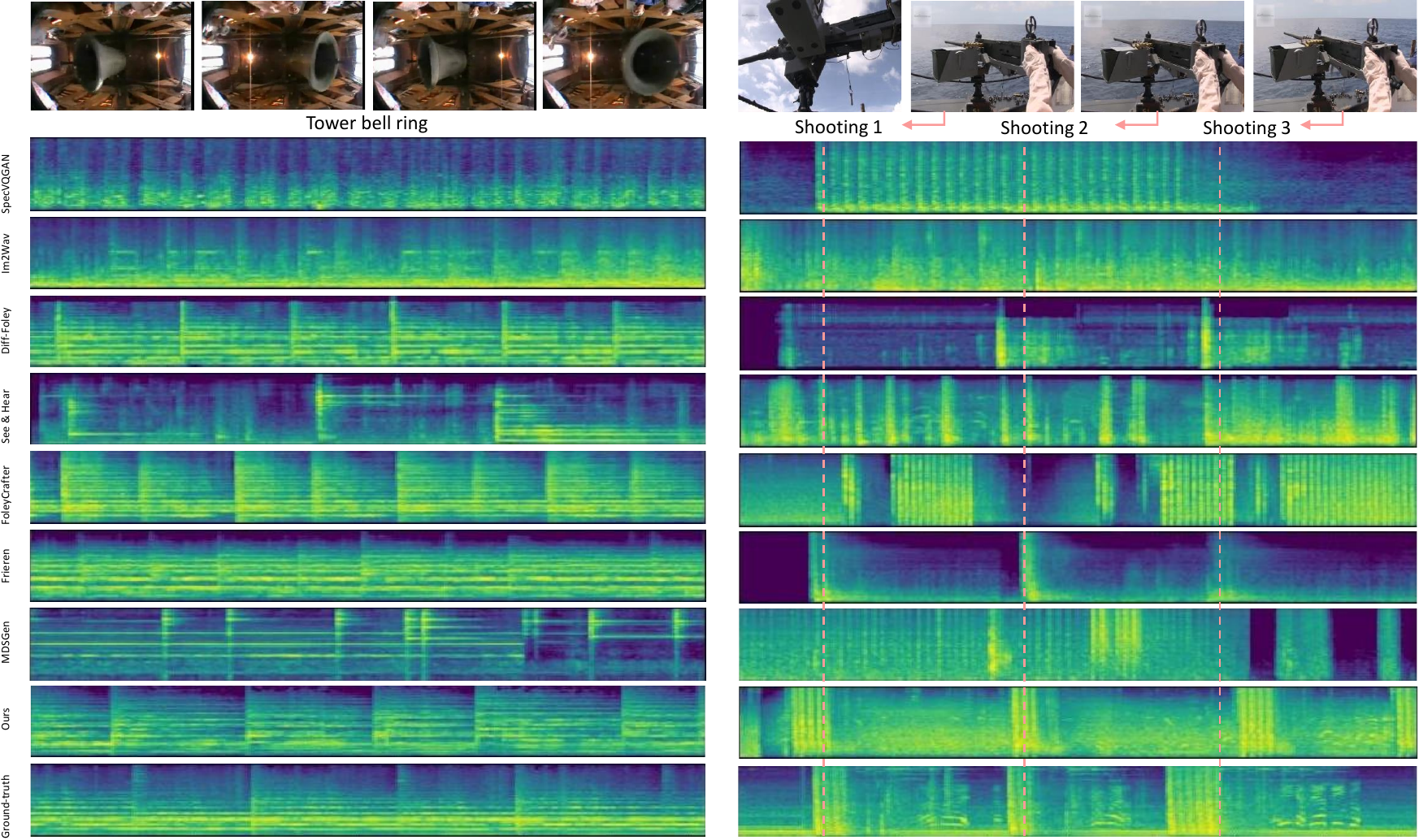}
    \caption{\textbf{Visual comparisons on \textit{6vl7eSBL-ag\_000090} and \textit{OFOHNgpDS38\_000091} from the VGGSound dataset.}}
    \label{fig:sup_3}
\end{figure*}

Figs. \ref{fig:sup_1}, \ref{fig:sup_2}, and \ref{fig:sup_3} present spectrogram visualizations compare our method with prior works \cite{iashin2021taming, sheffer2023hear, luo2024diff, xing2024seeing, zhang2024foleycrafter, wang2025frieren, pham2024mdsgen}.
These visualizations illustrate how different models capture event-driven acoustic cues and maintain synchronization with visual content. Our method demonstrates more precise temporal alignment, clearer event transitions, and improved spectral fidelity compared to baselines, which often exhibit artifacts, misaligned sound events, or missing key audio cues. These results further validate the effectiveness of our proposed framework in generating high-quality, temporally coherent audio.

\end{document}